# Ponderomotive Instability of RF Cavities with Vector Sum Control, and Cure By Difference Control

S.R. Koscielniak*, TRIUMF*, Vancouver, B.C., Canada

*Abstract*

The radiation pressure inside an RF cavity resonator, resulting from the Lorentz force, couples the fundamental electromagnetic mode to the mechanical modes and may lead to static and dynamic ponderomotive instability. The analysis for generator driven (GD) cavity and for self-excited (SEL) cavity was given by Schulze in 1971 and by Delayen in 1978, respectively. Since that time, little or no new analytical work - until now. Here are presented results from the analysis for multiple (N) cavities (GD or SE) powered from a single source and regulated in vector sum. The system behaves as N virtual cavities, one with all loops present and N-1 with no loops except the tuner. After reviewing stability threshold criteria for these systems, difference control in the tuning loops is proposed as means for extending the stability range. Counter-tuning of alternating cavities is also discussed.

## INTRODUCTION

Sketches of the Generator Driven (GD) and Self-Excited Loop (SEL) configurations of RF resonator control are given by Schilcher[1] chapter 2.3.1. But the sketches omit the resonator tuning control (applied externally to the cavity) and the internal detuning arising from Lorentz forces and the tuning disturbances arising from microphonics – all of which are a key feature of this paper.

The original ponderomotive stability analysis for GD was given by Schulze [2] in 1971, and for SEL by Delayen [3] in 1978. Schulze treated the case of resonator and no loops by a variety of methods including the Routh-Hurwitz (RH) analysis leading to algebraic stability thresholds. Schulze also studied the effect of feedback loops.

Inspired by an oscillatory instability observed [4] at the ARIEL 2-cavity cryomodule, we build on the works of Schulze and Delayen to the case of multiple cavities driven from a single RF source. Results for GD are presented here. Most GD results have SEL analogues. Indeed, any GD result that relies on a "symmetry argument" carries over to SEL case. However, space does not permit to report SE results here. Interested reader should consult the references [5-8] to IPAC'19. All results have been derived analytically (by Routh-Hurwitz analysis in *Mathematica*[9]). However, many results presented here are numerical examples – to avoid writing long equations.

## *Lorentz Force Detuning*

A RF cavity responds to internal radiation pressure by changing shape, resulting in volume changes ΔV. Slater's Theorem [10] gives the change in resonance frequency $\omega_c$. For example, Yamazaki [11] gives an expression for the detuning of a pill-box cavity. If the electric & magnetic field amplitudes (E &H) are time varying, then the Lorentz force detuning is time dependent; and includes dynamics of cavity-wall inertia and elasticity.

## *Cavity Mechanical & Electrical Response*

Coupling coefficients between EM frequency shift and cavity voltage are defined at DC. Let $\Delta\omega_c = -k_{DC} V_0^2$ where $k_{DC} = \Sigma_m k_m$ is the sum of mechanical modes. Mode *m* response to (DC) voltage modulations ($a_c$): $\delta\omega_c = -2k_m V_0^2 a_v$. Let – $\delta\omega_c \tau_c \equiv K_m a_v$ and $K_L = \Sigma_m K_m$ be normalized values. The mechanical resonator response, MM[s], extends dynamical behavior to AC.

The Cavity EM fundamental mode is modelled by a parallel resonance LCR circuit. Consider a pure sinusoidal oscillation Exp[+jωt] where ω is the drive frequency. The drive current source is $I_g$. The response voltage is $V = Z \times I_g$ where impedance Z=RCos[ψ]Exp[+jψ] and detuning angle (ψ) quantifies the difference between drive frequency ω and resonance frequency $\omega_c$ of the cavity. Quantities are

$$\text{Tan}[\Psi]=(\omega_c^2-\omega^2)/(2\alpha\omega) \text{ and } V_0 = V_g\text{Cos}[\Psi]$$

and $\alpha=\omega_c/(2Q_c)$ and $\tau_c = 1/\alpha$ is cavity time constant = 2/(cavity EM bandwidth).

## PONDEROMOTIVE INSTABILITY

## *Monotonic instability*

We have enough information to introduce a fundamental GD stability condition: $\Psi < 0 \rightarrow \omega$ drive> ω resonance; which is a limitation to and by the Lorentz force detuning.

Near DC, changes of frequency and voltage depend on the local derivative of the resonance curve.
MM Resonator Response: $(\partial\omega/\partial a_V) \approx -2K_L\text{Cos}[\psi]$
EM Resonator Response: $(\partial a_V/\partial\omega) \approx -\text{Sin}[\psi]$
At threshold, the product of amplification factors is unity.
$(\partial\omega/\partial aV)^{MM}(\partial aV/\partial\omega)^{EM} \approx 2K_L\text{Cos}[\psi]\text{Sin}[\psi] =1$
Hence the threshold for monotonic instability:

$$K_L < 1/(2\text{Cos}\psi\ \text{Sin}\psi) = 1/\text{Sin}[2\psi].$$

$K_L \propto V_0^2$ so eventually the instability is always reached. The instability is near DC because this is the only frequency at which mechanical modes can all cooperate. The monotonic threshold is insensitive to dynamics.

## *Oscillatory instability*

The EM resonator pumps the mechanical mode (MM) displacement amplitude; while the MM pumps the voltage modulation index $a_v = \delta V/V_0$ – but not $V_0$. Because the single MM oscillates at (or near) its resonance frequency, $\Omega_m$, its response is boosted by MM quality factor $Q_m$. Hence $\delta\omega_c \tau_c \approx -2Q_m K_m a_v$. During mechanical oscillation, the cavity EM resonance moves up and down in frequency. The effect is to drive the cavity at upper and lower sidebands

---

*TRIUMF receives funding via a contribution agreement with the National Research Council of Canada
† shane@triumf.ca

($\omega \pm \Omega_m$) with FM depth $\Omega_m$, leading to changes in the amplitude response. Differencing of sidebands leads to a net excitation $\propto \cos(\psi+\delta\psi) - \cos(\psi-\delta\psi)$ with $\delta\psi \approx \Omega_m\tau_c$, leading to $a_v \approx 2\sin\Psi\cos\Psi\,(\delta\omega_c\tau_c)(\Omega_m\tau_c)$. At threshold, the amplification factor is unity. Hence oscillatory threshold:

$K_m \approx -(1+\rho^2)/[2\rho Q_m K_m \sin(2\Psi)]$ where $\rho = \Omega_m\tau_c$.

This particular expression for $K_m$ is valid for $\rho \approx 1$. The oscillatory instability occurs above resonance (i.e. $\Psi<0$); and threshold depends on dynamical parameter $\rho = \tau_c \Omega_m$ = (EM time constant)×(MM frequency). $\rho$ answers the question: is $\Omega_m$ inside or outside cavity EM bandwidth? $\rho$ has large dynamic range: $\rho \gg 1$ light loading of cavity $Q_c$ (see references), typically very stable; $\rho \approx 1$ heavy loaded $Q_c$ regime (example here), typically prone to instability. Mechanical Mode dynamics is governed by:

$$MM[s] \to \frac{Q_m \Omega_m^2}{s\,\Omega_m + Q_m(s^2 + \Omega_m^2)}$$

*Note Bene*: these ponderomotive instabilities have some similarity to the Robinson instability [12] in synchrotron or storage rings. MM takes the role of charged particle beam.

### Sensitivity of thresholds

For mathematical analysis, the natural and simplest detuning variable is angle $\Psi$. But in the real world we deal with angular frequency $\omega$. Therefore, we are interested how the threshold varies with $\omega$.

$\partial K/\partial \omega = (\partial K/\partial \Psi)(\partial \Psi/\partial \omega) \approx (\partial K/\partial \Psi)\cdot(-2Q_c/\omega_c)\cos^2\Psi$. Hence the sensitivity is greater for small cavity bandwidth $\omega_c/Q_c$ (i.e. high $Q_c$). To get a real feeling for how very sensitive is the threshold with respect to errors in cavity tuning, consider the following: in the range $\Psi = [0, \pi/4]$, which maps to the tuning range $\Delta\omega = [0, \tfrac{1}{2}$ bandwidth $]$, the threshold varies from infinite to minimum value $K_L = 1$.

A "microphonic" is an excitation of a mechanical mode. A subset of the modes couple to the EM resonance frequency; and such excited modes will eat into the stable detuning region and stability margin.

## INSTABILITY CLASSIFICATION

The analysis contains the following steps: (1) Find steady state solution of nonlinear dynamical equation for RF cavity coupled to linear mechanical resonator; (2) Make small perturbations & discard products of small quantities; (3) Laplace transform (convert the ODE to an algebraic equation); (4) Obtain the characteristic equation; and (5) Apply Routh-Hurwitz criteria, or find roots numerically.
Instability classification as follows:
- Complex frequency s
- Monotonic instability: $Re[s] \geq 0$ and $Im[s]=0$
   - threshold depends on $K_L$ & $\psi$
- Oscillatory instability: $Re[s] \geq 0$ and $Im[s] \neq 0$
   - threshold depends on $K_m$, $Q_m$, $\rho$ & $\psi$

### Single Cavity – no loops

The system equations are at the head of next column:

$$\begin{pmatrix} a_v(1+s\tau_c) + p_v \tan[\Psi] \\ -\delta\omega_m \tau_c + p_v(1+s\tau_c) - a_v \tan[\Psi] \\ 2\,MM[s]\,a_v K_L + \delta\omega_m \tau_c \end{pmatrix} == 0$$

The characteristic polynomial is:

$$(1+s\tau_c)^2 - 2\,MM[s]\,K_L \tan[\Psi] + \tan[\Psi]^2$$

Monotonic threshold: $K_L < 1/\sin[2\psi]$.
Oscillatory threshold $\propto \rho/Q_m$ which may be less than $K_m$. The thresholds are plotted in Fig.1.

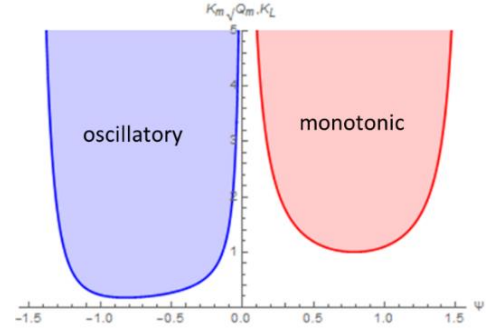

Figure 1: Instability thresholds $K_m$ and $K_L$ for cavity with no control loops. Other microphonics rattle the frequency/tuning origin reducing the stable working area.

### Single Cavity – with only fast tuning control

The system equations are:

$$\begin{pmatrix} a_v(1+s\tau_c) + p_v \tan[\Psi] \\ -\delta\omega_m \tau_c - \delta\omega_T \tau_c + p_v(1+s\tau_c) - a_v \tan[\Psi] \\ K_t\,p_v + \delta\omega_T \tau_c \\ 2\,MM[s]\,a_v K_L + \delta\omega_m \tau_c \end{pmatrix} == 0$$

The characteristic polynomial is:

$$(1+s\tau_c)(K_t + s\tau_c) - 2\,MM[s]\,K_L \tan[\Psi] + \tan[\Psi]^2$$

Monotonic threshold $K_L < 1/2\,(K_t \cot[\psi] + \tan[\psi])$
Oscillatory threshold $K_m \propto K_t/Q_m$ when $\rho \approx 1$;
Need minimum gain of $K_t > \sqrt{Q_m}$ when $\rho \approx 1$
The thresholds are plotted in Fig.2.
If tuner time constant $> \tau_c$ then get additional instability.

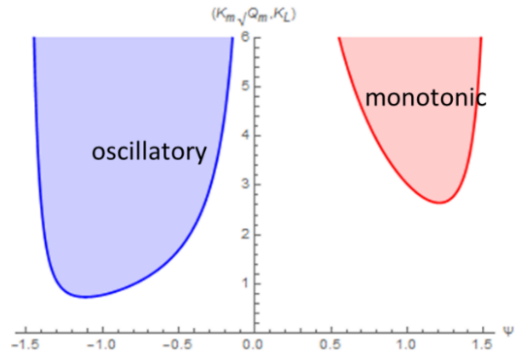

Figure 2: Instability thresholds $K_m$ and $K_L$ for cavity with only (fast) tuning control. $Q_m = 50$ and $K_t = 7$.

### Single Cavity – all control loops

Proportional control gains $K_a$, $K_p$, $K_t$ (amplitude, phase, tuning). The system equations (equal to 0) are:

$$\begin{pmatrix} -a_g + a_v(1+s\,\tau_c) - p_g \tan[\Psi] + p_v \tan[\Psi] \\ -p_g - \delta\omega_m \tau_c - \delta\omega_T \tau_c + p_v(1+s\,\tau_c) + a_g \tan[\Psi] - a_v \tan[\Psi] \\ a_g + a_v K_a \\ p_g + K_p p_v \\ -K_t p_g + K_t p_v + \delta\omega_T \tau_c \\ 2\,MM[s]\,a_v K_L + \delta\omega_m \tau_c \end{pmatrix}$$

The characteristic polynomial is:

$$(K_a + s\,\tau_c)(K_p K_t + s\,\tau_c) + K_p \tan[\Psi]\left(-2\,MM[s]\,K_L + K_a \tan[\Psi]\right)$$

Monotonic threshold boosted by $K_a$, $K_t$ (not $K_p$):

$$K_L < 1/2\, K_a (K_t \cot[\psi] + \tan[\psi])$$

Oscillatory threshold is also raised: when $\rho \approx 1$ need gain product larger than Q mechanical: $K_a \times K_p \times K_t > Q_m$.
The thresholds are plotted in Fig.3.

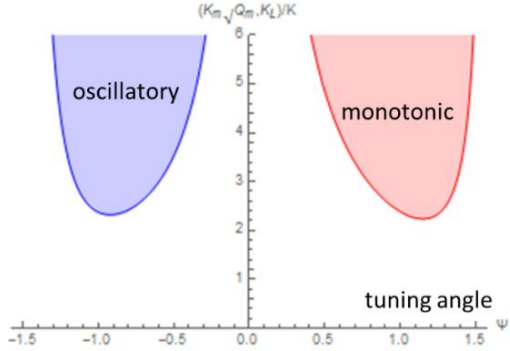

Figure 3: Instability thresholds $K_m$ and $K_L$ for cavity with all control lops. $Q_m = 100$ and $K_a = K_p = K_t = 5$.

# VECTOR SUM CONTROL

## Introduction

Sum cavity voltages together respecting amplitudes and phases to get a net phasor and apply to the input of a single AM/PM (or I/Q) feedback loop to the shared RF source. The motivation for vector sum control [13-15] is the shared RF source. High power RF sources are expensive, and so it is more cost effective to drive several cavities from one source. Typically, a cryomodule contains a cavity string; and we attempt to build cavities with identical EM modes. Thus, the mechanical modes will also be very similar. This leads to equations that are equal/symmetrical in MM[s]. We shall assume the vector sum control is supplemented with simple tuning control of each cavity; if the tuning control is omitted, then the instability thresholds will be lower. We demonstrate the approach using the simplest case of two RF cavities.

## Case A: Two Cavities in Vector Sum with Simple Tuning Control

The cavities are considered identical, they have equal detunings, and the control is applied symmetrically leading to the system equations (equal to 0) given at the head of the next column.

The characteristic equation factors into the two polynomials given above; and from the control viewpoint, behaves exactly like 2 virtual cavities:
 a) Cavity with all control loops present
 b) Cavity with no loops except tuner.

$$\begin{pmatrix} -a_g + (1+s\,\tau_c)\,a_{v,1} - p_g \tan[\Psi] + p_{v,1} \tan[\Psi] \\ -p_g + (1+s\,\tau_c)\,p_{v,1} - \tau_c \delta\omega_{m,1} - \tau_c \delta\omega_{T,1} + a_g \tan[\Psi] - a_{v,1} \tan[\Psi] \\ -\tfrac{1}{2} K_t p_g + \tfrac{1}{2} K_t p_{v,1} + \tfrac{1}{2} \tau_c \delta\omega_{T,1} \\ 2\,MM[s]\,K_L a_{v,1} + \tau_c \delta\omega_{m,1} \\ a_g + \tfrac{1}{2} K_a a_{v,1} + \tfrac{1}{2} K_a a_{v,2} \\ p_g + \tfrac{1}{2} K_p p_{v,1} + \tfrac{1}{2} K_p p_{v,2} \\ -a_g + (1+s\,\tau_c)\,a_{v,2} - p_g \tan[\Psi] + p_{v,2} \tan[\Psi] \\ -p_g + (1+s\,\tau_c)\,p_{v,2} - \tau_c \delta\omega_{m,2} - \tau_c \delta\omega_{T,2} + a_g \tan[\Psi] - a_{v,2} \tan[\Psi] \\ -\tfrac{1}{2} K_t p_g + \tfrac{1}{2} K_t p_{v,2} + \tfrac{1}{2} \tau_c \delta\omega_{T,2} \\ 2\,MM[s]\,K_L a_{v,2} + \tau_c \delta\omega_{m,2} \end{pmatrix}$$

Virtual cavity "b" goes unstable before cavity "a". This follows from symmetry; and is therefore also true for SEL. Consequence: no matter how hard we push the gains $K_a$ and $K_p$ there is no improvement of ponderomotive stability - because virtual cavity "b" is the problem.

## Sum and Difference Control

Introduce the vectors

$\mathbf{V_{sum}} = \{a_1+a_2, \varphi_1+\varphi_2\}$ and $\mathbf{V_{diff}} = \{a_1 - a_2, \varphi_1 - \varphi_2\}$.

And likewise for all other variables in the state vector. Written in terms of these new variables, we find [16] that the system equations divide into two separate sets. $\mathbf{V_{sum}}$ is governed by virtual cavity "a" (all loops present); whereas $\mathbf{V_{diff}}$ is governed by virtual cavity "b" (tuning loop only). This implies that the difference mode will be the first to go unstable. We conclude that vector sum control is ineffective for the $V_1 - V_2$ state variable; and this will lead to a lowered threshold for ponderomotive instability as compared to a single cavity with amplitude, phase and tuning loops present.

Restoring control over the cavity voltages, implies introducing control that is different between the cavities; and (when there is a single RF source) the only place to do this is at the (fast) cavity tuners

Raising or removing ponderomotive thresholds implies targeted DC-coupled feedback for the monotonic instability; and targeted AC-coupled feedback for the oscillatory instability. Recall the damped harmonic oscillator: $x'' + kx' + \Omega^2 = 0$. For damping, need a quadrature term like $kx'$; and so take: $\delta\omega_1 = +k\,s\,\Delta\varphi$ and $\delta\omega_2 = -k\,s\,\Delta\varphi$ where $\Delta\varphi = (\varphi_1 - \varphi_2)$ where the constant k is to be determined.

## Cases B & C: Two Cavities in Vector Sum with Difference & Derivative Controls

Motivated as above, consider two further cases.
 A) Retain sum control for amplitude and phase loops; and introduce difference control for tuning loops. The tuning control equations:

$$\begin{pmatrix} \tfrac{1}{2}(-K_t(p_g - p_{v,1} + p_{v,2}) + \tau_c \delta\omega_{T,1}) \\ \tfrac{1}{2}(-K_t(p_g + p_{v,1} - p_{v,2}) + \tau_c \delta\omega_{T,2}) \end{pmatrix}$$

 B) Retain controls as for case B and add derivative control to the tuning loops. The control equations:

$$\begin{pmatrix} \tfrac{1}{2}(-(1+s\,T)\,K_t(p_g - p_{v,1} + p_{v,2}) + \tau_c \delta\omega_{T,1}) \\ \tfrac{1}{2}(-(1+s\,T)\,K_t(p_g + p_{v,1} - p_{v,2}) + \tau_c \delta\omega_{T,2}) \end{pmatrix}$$

The following figures 4-11 compare the three cases A, B, C and, for good measure, the case Z of vector sum but no tuning control, expected to have the lowest threshold.

Figures.4-7 show the root locus of the growth rate, Re[s], as the collective coupling coefficients $K_L$ is progressively increased for positive detuning, $\Psi>0$, leading to monotonic instability. Figures. 8-11 show the root locus of the growth rate, Re[s], as the single mode coupling coefficient $K_m$ is progressively increased for negative detuning, $\Psi<0$, leading to oscillatory instability.

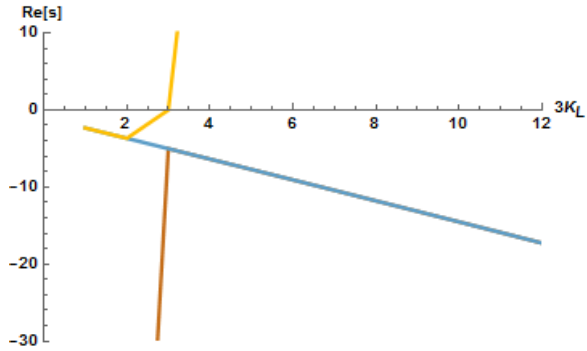

Figure 4: Case Z: $Q_m = 100$, $Ka = Kp = 5$, $Kt = 0$, $\Psi>0$

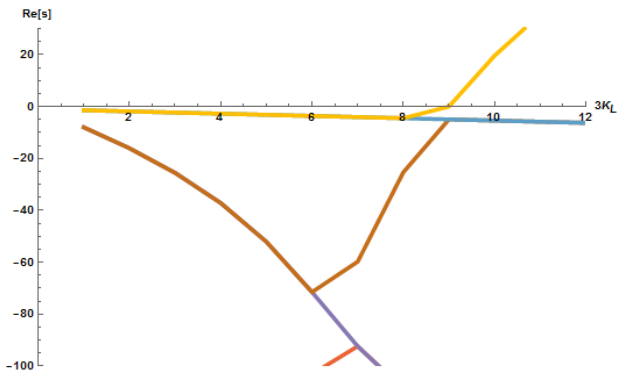

Figure 5: Case A: $Q_m = 100$ and $Ka = Kp = Kt = 5$, $\Psi>0$

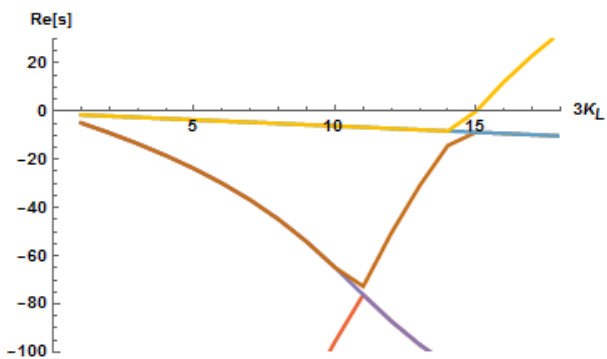

Figure 6: Case B: tuning with difference variables, $\Psi>0$

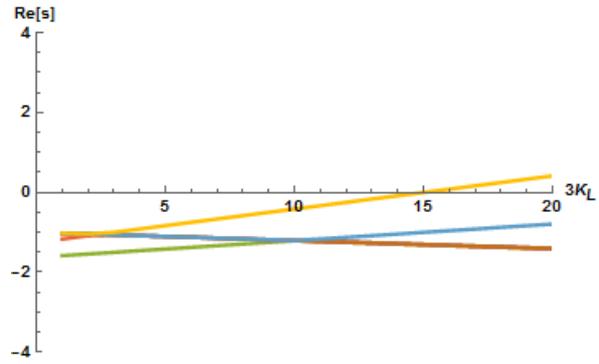

Figure 7: Case C: differential control added, $\Psi>0$

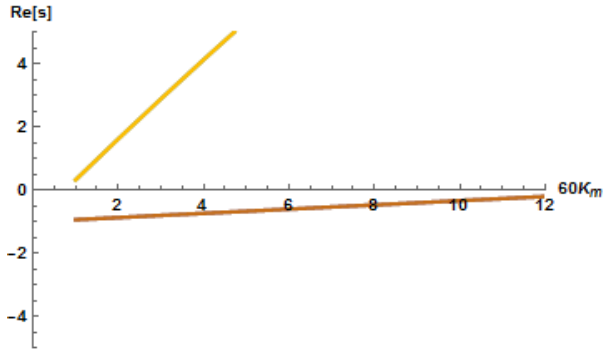

Figure 8: Case Z: $Q_m = 100$, $Ka = Kp = 5$, $Kt = 0$, $\Psi<0$.

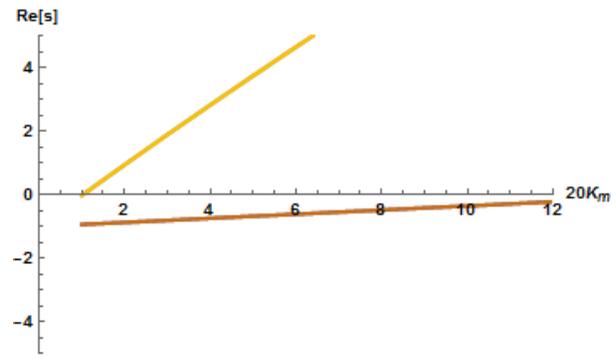

Figure 9: Case A: $Q_m = 100$ and $Ka = Kp = Kt = 5$, $\Psi<0$

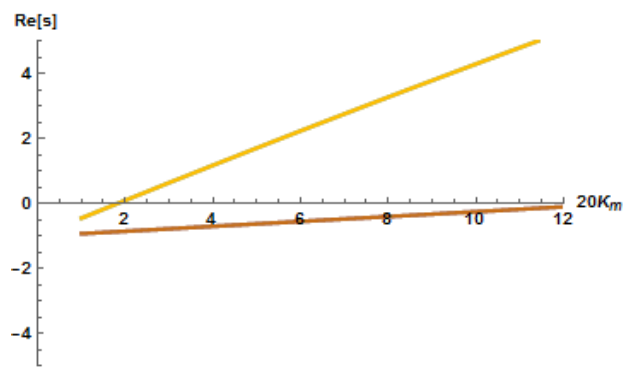

Figure 10: Case B: tuning with difference variables, $\Psi<0$

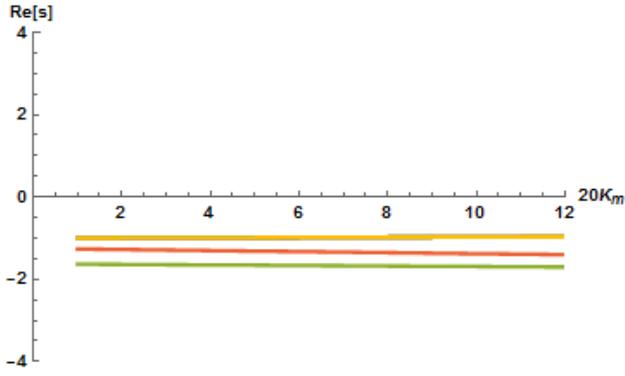

Figure 11: Case C: differential control added, $\Psi<0$.

# N-CAVITY CONTROL

For N cavity resonators with a shared single RF source and vector sum control, the characteristic equation factors into N polynomials; and behaves exactly as:
- 1 virtual cavity with all loops present
- N-1 virtual cavities with no control loops except a tuner for each.

This implies we can employ the stability criteria for single cavities; and all the ponderomotive difference modes have lower threshold than the sum mode.

In short pulse operation, the instability is no concern. The initial values are periodically reset before the instability has time to grow significantly. However, for long pulse, near c.w. or c.w. operation the instabilities can become manifest. Hence we need to add difference control. Above its threshold, simple tuning control cannot straighten the vector sum. Without changing the cavity $Q_c$ or access to the individual modulations $(a_i,p_i)$ sum & difference is the only way to raise the threshold or damp the difference modes.

## Difference Control/Variables

We have to generalize the concept of differences [16] to extend the tuning control from two to N cavities. Treating the variables symmetrically avoids breaking the degeneracy of N-1 characteristics. The 3-cavity system will demonstrate the principle: cyclic permutation.

$$\begin{pmatrix} p_{v,1} - p_{v,2} - p_{v,3} \\ -p_{v,1} + p_{v,2} - p_{v,3} \\ -p_{v,1} - p_{v,2} + p_{v,3} \end{pmatrix}$$

We choose to incorporate some flexibility, through a coefficient $c$, to be determined. Hence the system equations (equal to 0) are those given at the head of the next column.

From these equations we can proceed to the N characteristics and the stability analysis. Or we can make a detour (helpful, but not essential) and rewrite the equations in simpler form in terms of new variables.

## Decoupling Variables

The amplitude and phase and detuning variables (dimensionless modulation indices), a set for each RF resonator, are the usual coordinates. "Difference variables" are mixtures of the phase coordinates that optimize the detuning control so as to maximize the ponderomotive stability.

$$\begin{pmatrix} -a_g + (1 + s\,\tau_c)\,a_{v,1} + (-p_g + p_{v,1})\,\text{Tan}[\Psi] \\ -p_g + (1 + s\,\tau_c)\,p_{v,1} - \tau_c\,\delta\omega_{m,1} - \tau_c\,\delta\omega_{T,1} + a_g\,\text{Tan}[\Psi] - a_{v,1}\,\text{Tan}[\Psi] \\ \frac{1}{3}\left(\tau_c\,\delta\omega_{T,1} - (p_g - p_{v,1} + c\,(p_{v,2} + p_{v,3}))\,K_{\tau_c}[s]\right) \\ 2\,K_m\,MM[s]\,a_{v,1} + \tau_c\,\delta\omega_{m,1} \\ -a_g + (1 + s\,\tau_c)\,a_{v,2} + (-p_g + p_{v,2})\,\text{Tan}[\Psi] \\ -p_g + (1 + s\,\tau_c)\,p_{v,2} - \tau_c\,\delta\omega_{m,2} - \tau_c\,\delta\omega_{T,2} + a_g\,\text{Tan}[\Psi] - a_{v,2}\,\text{Tan}[\Psi] \\ \frac{1}{3}\left(\tau_c\,\delta\omega_{T,2} - (p_g - p_{v,2} + c\,(p_{v,1} + p_{v,3}))\,K_{\tau_c}[s]\right) \\ 2\,K_m\,MM[s]\,a_{v,2} + \tau_c\,\delta\omega_{m,2} \\ -a_g + (1 + s\,\tau_c)\,a_{v,3} + (-p_g + p_{v,3})\,\text{Tan}[\Psi] \\ -p_g + (1 + s\,\tau_c)\,p_{v,3} - \tau_c\,\delta\omega_{T,3} - \delta t\,\omega_{m,3} + a_g\,\text{Tan}[\Psi] - a_{v,3}\,\text{Tan}[\Psi] \\ \frac{1}{3}\left(\tau_c\,\delta\omega_{T,3} - (p_g + c\,(p_{v,1} + p_{v,2}) - p_{v,3})\,K_{\tau_c}[s]\right) \\ 2\,K_m\,MM[s]\,a_{v,3} + \delta t\,\omega_{m,3} \\ a_g + \frac{1}{3}(a_{v,1} + a_{v,2} + a_{v,3})\,K_a[s] \\ p_g + \frac{1}{3}(p_{v,1} + p_{v,2} + p_{v,3})\,K_p[s] \end{pmatrix}$$

"Decoupling variables" are mixtures of the usual coordinates that take the system matrix into block diagonal form (with fewer, simpler coupling elements) revealing the underlying structure/causation/nature of the system.

Decoupling variables are an aid to understanding and comprehension of the system stability but are not necessarily a basis for control.

For NC = 2 RF cavities, sum-and-difference variables are synonymous with decoupling variables. For NC > 2 cavities, sum and difference variables are NOT the decoupling variables.

## Change of Vector Basis

The characteristic is the determinant of the system matrix. From linear algebra theory, we know the characteristic is independent of basis vector. But the appearance of the system matrix **P** depends on the choice of basis vector. So, underlying symmetry can be made manifest by suitable choice of basis. Further, the ease or difficulty of computing the determinant may change. For large N, say N>4, computations are greatly facilitated by block diagonal form.

Any linear superposition of the old bases is also a basis. The transform to decoupling variables is generated by a matrix **T**. [If **T** can be inverted then old and new basis vectors are independent.] Old basis vector **v**; new basis vector **v'**=**Tv**. The new system matrix **P'**=**TPT**$^{-1}$ is given by a similarity transform. The 3-cavity system will demonstrate the principle. The transformation matrix is **T**=

$$\begin{pmatrix} 0 & 0 & 0 & 0 & 0 & 0 & 0 & 0 & 0 & 0 & 0 & 1 & 0 \\ 0 & 0 & 0 & 0 & 0 & 0 & 0 & 0 & 0 & 0 & 0 & 0 & 1 \\ \frac{1}{3} & 0 & 0 & 0 & \frac{1}{3} & 0 & 0 & 0 & \frac{1}{3} & 0 & 0 & 0 & 0 \\ 0 & \frac{1}{3} & 0 & 0 & 0 & \frac{1}{3} & 0 & 0 & 0 & \frac{1}{3} & 0 & 0 & 0 \\ 0 & 0 & \frac{1}{3} & 0 & 0 & 0 & \frac{1}{3} & 0 & 0 & 0 & \frac{1}{3} & 0 & 0 \\ 0 & 0 & 0 & \frac{1}{3} & 0 & 0 & 0 & \frac{1}{3} & 0 & 0 & 0 & \frac{1}{3} & 0 & 0 \\ 1 & 0 & 0 & 0 & -1 & 0 & 0 & 0 & 0 & 0 & 0 & 0 & 0 \\ 0 & 1 & 0 & 0 & 0 & -1 & 0 & 0 & 0 & 0 & 0 & 0 & 0 \\ 0 & 0 & 1 & 0 & 0 & 0 & -1 & 0 & 0 & 0 & 0 & 0 & 0 \\ 0 & 0 & 0 & 1 & 0 & 0 & 0 & -1 & 0 & 0 & 0 & 0 & 0 \\ 1 & 0 & 0 & 0 & 0 & 0 & 0 & 0 & -1 & 0 & 0 & 0 & 0 \\ 0 & 1 & 0 & 0 & 0 & 0 & 0 & 0 & 0 & -1 & 0 & 0 & 0 \\ 0 & 0 & 1 & 0 & 0 & 0 & 0 & 0 & 0 & 0 & -1 & 0 & 0 \\ 0 & 0 & 0 & 1 & 0 & 0 & 0 & 0 & 0 & 0 & 0 & -1 & 0 & 0 \end{pmatrix}$$

The relation between new and old variables is at the head of the next column:

$$\begin{pmatrix} a_g \\ p_g \\ A_1 \\ P_1 \\ \tau O_1 \\ \tau M_1 \\ A_2 \\ P_2 \\ \tau O_2 \\ \tau M_2 \\ A_3 \\ P_3 \\ \tau O_3 \\ \tau M_3 \end{pmatrix} == \begin{pmatrix} a_g \\ p_g \\ \frac{1}{3}(a_{v,1} + a_{v,2} + a_{v,3}) \\ \frac{1}{3}(p_{v,1} + p_{v,2} + p_{v,3}) \\ \frac{1}{3}\tau_c (\delta\omega_{T,1} + \delta\omega_{T,2} + \delta\omega_{T,3}) \\ \frac{1}{3}\tau_c (\delta\omega_{m,1} + \delta\omega_{m,2} + \delta\omega_{m,3}) \\ a_{v,1} - a_{v,2} \\ p_{v,1} - p_{v,2} \\ \tau_c (\delta\omega_{T,1} - \delta\omega_{T,2}) \\ \tau_c (\delta\omega_{m,1} - \delta\omega_{m,2}) \\ a_{v,1} - a_{v,3} \\ p_{v,1} - p_{v,3} \\ \tau_c (\delta\omega_{T,1} - \delta\omega_{T,3}) \\ \tau_c (\delta\omega_{m,1} - \delta\omega_{m,3}) \end{pmatrix}$$

Quantities on the right are sum & difference modes. The new system matrix **P'** is block diagonal:

### N-Cavity System Stability

The determinant can be read off from the matrix blocks:

$\text{Det}[P'] = (1/3^3) \cdot \text{poly1}^2 \cdot \text{poly2}$ where the polynomials are:

```
poly1 == (1 + s τ_c) (c (-1 + K_t) + K_t + s τ_c)
        - 2 Km MM[s] Tan[Ψ] + Tan[Ψ]^2
poly2 == (K_a + s τ_c) (-2 c (-1 + K_t) + K_p K_t + s τ_c)
        - 2 Km MM[s] K_p Tan[Ψ] + K_a K_p Tan[Ψ]^2
```

[For brevity, here $K_a$, $K_p$, $K_t$ stand in place of $1+K_a$, $1+K_p$, $1+K_t$.] Coefficient $c>0$ will increase/decrease the stability of virtual cavity b/a respectively. For example, $c=1$ will double the monotonic threshold (of poly1). Note, for N=3 there are two virtual cavity b, and one virtual cavity a.

For general N, the determinant is

$$\text{Det}[P] = (1/N^N) \cdot \text{poly1}^{(N-1)} \cdot \text{poly2} .$$

This form maximizes the stability. If we had taken a less symmetric set of difference variables (as the basis of tuning control) then the factorization leads to at least one polynomial with roots at smaller $K_L$ or $K_m$ than poly1. As N is increased, the virtual cavity "a" (poly2) has to stabilize progressively more virtual cavities "b" (poly1). Hence, we shall expect some limitations.

The condition that the monotonic threshold of cavity "a" falls below that of cavity "b" is disastrous and must be avoided. The upper limit on $c$ is:

$$c == \frac{(-1 + K_a) K_p (K_t + \text{Tan}[\Psi]^2)}{((-1 + N) K_a + K_p) (-1 + K_t)}$$

The limit scales inversely as the number of cavities (1/N) in the string powered from a single source, and can be raised by increasing the phase gain $K_p$. Other limits on $c$ arise from stipulating desired/acceptable improvements/ reductions of the cavity "a" / "b" thresholds, respectively; but the condition above is fundamental.

### OPPOSITE CAVITY DETUNING

We return to the case of two cavities only, and take a hint from alternating gradient focusing. Consider cavities with equal and opposite detuning, and sharing a vector sum control. We might expect that cavity #1, Ψ<0, oscillatory instability is reduced by cavity #2; while cavity #2, Ψ>0, monotonic instability is reduced by cavity #1.

The system equations are the same as Case A above, except one cavity has +Ψ → -Ψ. With opposite detunings the characteristic does not factor, leading to an octic equation. The polynomial coefficients contain only powers of $(\text{Tan}[\Psi])^2$; so the characteristic has identical roots at ±Ψ. The result is 1 virtual cavity that behaves the same at ±Ψ. There are equal monotonic thresholds at ±Ψ; and equal oscillatory thresholds at ±Ψ. Both thresholds are higher than for single cavity with tuning loop alone. Both thresholds are lower than for single cavity with all loops present. Hence "opposite detuning" wins because it is more stable than a virtual cavity with tuning loop alone. See Figs 12-14. For example, the monotonic threshold

$$K_L < 1/2 \sqrt{K_a} (K_t \text{Cot}[\psi] + \text{Tan}[\psi])$$

falls between those of virtual cavities "b" and "a".

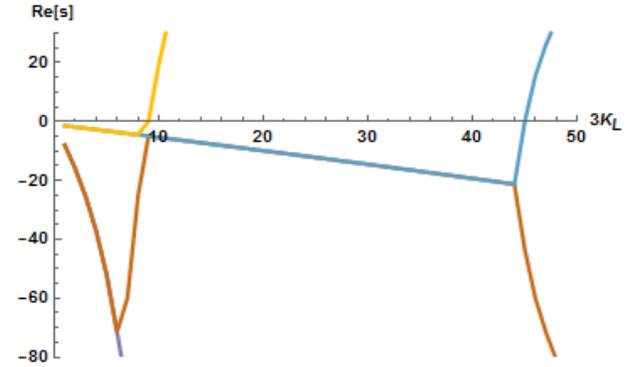

Figure 12: Case A: monotonic, Ψ>0. 1st threshold (left) is tuner only; 2nd threshold (right) all loops present.

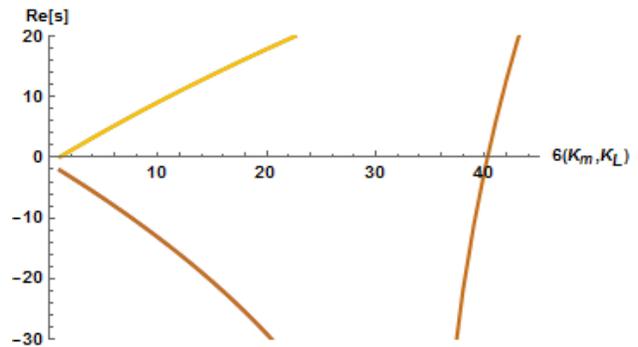

Figure 13: Opposite detunings: oscillatory (left) and monotonic (right) thresholds.

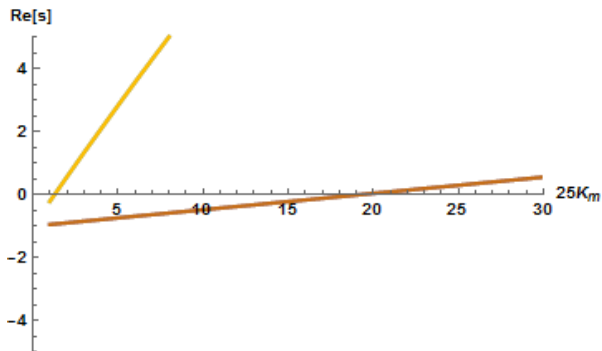

Figure 14: Case A: oscillatory, $\Psi<0$. 1st threshold (left) is tuner only; 2nd threshold (right) all loops present.

## CONCLUSIONS

$\rho = \tau_c \Omega_m$ is important parameter for oscillatory instability. The heavy loaded regime $\rho \approx 1$ is prone to oscillatory instability.

Multiple cavities driven in vector sum have an instability threshold equal that of a single cavity with only a tuning loop; and it does not matter how large are made the amplitude and phase loop gains.

Introducing difference control at the individual cavity tuners allows to straighten the vector sum; And to raise the monotonic threshold; And to eliminate the oscillatory instability if derivative feedback is used.

Extending the concept of difference variables, enables the technique to be applied to N cavities driven from one source; and this opens the way for use of vector-sum in c.w. and near-c.w. applications.

Those properties/results which depends onl on "symmetry arguments" apply equally well to SE loop.

Finally, we note "counter detuning" of alternating cavities as a means to raise thresholds without applying additional feedback.

*Caveat*

All preceding text assumes that each cavity has the same mechanical mode (or spectrum of modes). Contrastingly, if the cavities have different mechanical modes then the characteristic does not factor. In such case, pure numerical methods may have to be adopted. With respect to decoupling variables, the system matrix has sums of modes (strong coupling) inside the blocks and differences of modes (weak coupling) outside; but the matrix is still sparse. If the sets of mechanical modes are not too different from one another, it may be that the system behaviour is not vastly different from that reported here.

For the case of N=2, and one cavity is missing the $K_m$ mode, the situation becomes both more complicated and often more stable than the limits described here. We have performed analysis for such a case with two-cavities; but not reported here.

Truly non-linear effects arising from large amplitude oscillation of the cavity voltage are not treated here. A preliminary investigation for single cavity with no control loops given in Refs. [17, 18].